# In-space manufacturing of optical lenses: Fluidic Shaping aboard the International Space Station

Omer Luria[a], Mor Elgarisi[a], Eytan Stibbe[b,], Michael López-Alegría[c], Crissy Canerday[d], Alexey Razin[a], Sivan Perl[a], Valeri Frumkin[a,§], Jonathan Ericson[a], Khaled Gommed[a], Daniel Widerker[a], Israel Gabay[a], Ruslan Belikov[e], Edward Balaban[e,*] and Moran Bercovici[a,f *]

[a] Faculty of Mechanical Engineering, Technion – Israel Institute of Technology, Haifa, Israel
[b] The Rakia Mission, 3 Shadal St. Tel Aviv-Yafo, Israel
[c] Axiom Space, 1290 Hercules Ave, Houston, TX, USA
[d] NASA Marshall Space Flight Center, Huntsville, AL, USA
[e] NASA Ames Research Center, Moffett Blvd., Moffett Field, CA, USA
[f] Department of Materials, ETH Zürich, Switzerland
[§] Current affiliation: Department of Mechanical Engineering, Boston University, Boston, MA, USA

[*] Corresponding authors: edward.balaban@nasa.gov (EB), mberco@technion.ac.il (MB)

## Abstract

In-space manufacturing technologies are vital for enabling advanced space missions and addressing logistical limitations of space exploration. While additive manufacturing has progressed rapidly, it still falls short of delivering the ultra-smooth surfaces required for optical elements. Fluidic Shaping is a novel method that harnesses surface tension under microgravity to form optical components with exceptionally smooth surfaces. This study demonstrates the feasibility and potential of Fluidic Shaping as a method for manufacturing optical components in space through two experiments performed aboard the International Space Station (ISS) during the Ax-1 mission. The first experiment involved fabricating centimeter-scale polymer lenses, solidifying them via ultraviolet (UV) curing, and analyzing the resultant optics upon their return to Earth. While sub-nanometric surface smoothness was achieved, some polymer lenses displayed unexpected thermo-chemical deformations, indicating complex polymerization dynamics unique to the microgravity environment. In the second experiment, a large-scale, 172 mm diameter water lens was deployed, confirming Fluidic Shaping's scalability and demonstrating basic optical functionality through image analysis. These experiments collectively underline the technique's relevance for both small-scale optics and large-aperture applications. Our results highlight critical considerations for future research, including optimizing polymerization processes and refining liquid-handling methods to advance practical, in-space optical manufacturing capabilities.

# 1. Introduction

As humanity continues to explore our solar system and beyond, technologies for in-space manufacturing will be indispensable, as long-duration missions must be self-sufficient. Additive manufacturing is of particular interest for space applications owing to its high versatility and has been repeatedly demonstrated in recent years onboard the ISS to fabricate a wide range of mechanical objects. Additive manufacturing also offers the potential to overcome launch constraints by fabricating objects larger than the launcher's payload fairing limits.[1–6] Although additive manufacturing has notable advantages when relatively complex geometries are required, it cannot be applied to fabrication of macro-scale optics. Optical elements, such as lenses and mirrors, whether on small or large scale, require extremely tight manufacturing tolerances and exceptionally low surface roughness (typically on the scale of nanometers) specifications, which are out of reach for all existing additive manufacturing technologies.[7–9]

A notable example of the need for optical manufacturing in space is corrective eyewear. During long-duration spaceflight, astronauts experience spaceflight-associated neuro-ocular syndrome (SANS), reporting decreased near vision of up to 1.5 diopters, potentially appearing as early as three weeks in microgravity.[10–12] As no known prevention or treatment for SANS exists, this syndrome poses a key risk for deep space exploration. Consequently, on extended missions, corrective eyewear may be essential to maintaining operational readiness and the quality of life of the crew. Given the unpredictable nature of vision decline, developing the capability to fabricate corrective eyewear in space may be very important. Another example where in-space manufacturing of optics may be required is in the creation of extremely large space telescopes. To achieve high resolution and light collection abilities, telescopes should ideally be as large as possible, yet launch limitations cap their size to just a few meters. As a direct consequence, direct imaging of exoplanets – widely regarded as the "holy grail" of modern astronomy – remains an elusive goal.[13–16]

Surprisingly, although optical components are at the forefront of many cutting-edge technologies, optical manufacturing has undergone very little innovation over the past centuries. The standard methods for the fabrication of lenses are based on mechanical processes such as grinding and polishing, or molding.[17–20] These processes are based on heavy machinery and require significant energy and water resources, which are inherently incompatible with space environments.

Fluidic Shaping is a novel approach for the fabrication of optical components, such as lenses and mirrors, that leverages the natural surface tension of liquids under microgravity.[21] Under weightlessness, the shape of a liquid volume is prescribed solely by any solid boundaries it is in contact with, with its free surfaces characterized by constant mean curvatures. Injecting a liquid into a simple circular ring boundary will result in the shape of an optical lens with two spherical free surfaces. The radii of curvature of the surfaces (and thus the optical prescription of the lens) can then be changed simply by adding or removing liquid. The resulting surfaces are exceptionally smooth, naturally exhibiting sub-nanometric surface roughness. If the lens is made of a curable liquid (such as a photopolymer), it can be solidified into a rigid component. Figure 1 depicts the process of fabricating a lens using Fluidic Shaping. Owing to the lack of body forces in the system, the method is scale-invariant and can be used in the same fashion to create extremely large lenses, requiring only a larger bounding frame and larger liquid volumes. Frumkin and Bercovici[21] and Elgarisi et al.[22] have demonstrated the method on Earth by injecting the polymer into an immiscible immersion liquid of equal density to that of the polymer, thus achieving neutral buoyancy conditions. They have shown how different types of lenses (spherical, aspherical, and freeform) can be obtained by tuning the shape of the frame, the liquid lens volume, and the deviation from perfect neutral buoyancy.

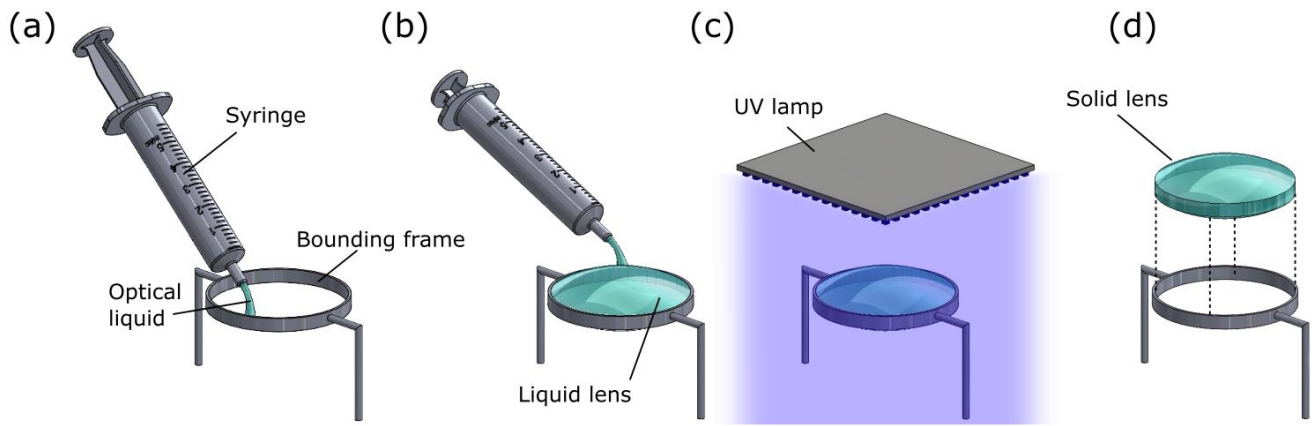

**Figure 1. Fabrication of an optical lens using Fluidic Shaping under microgravity.** (a) Optical liquid is injected into a circular bounding frame. (b) Once the frame is filled with liquid, the obtained minimum-energy state is a liquid lens with spherical free surfaces. (c-d) If the optical liquid is curable, it can be solidified (e.g., via UV light) to produce a solid lens.

Fluidic Shaping is inherently based on weightlessness conditions and is therefore directly applicable for in-space manufacturing of optics. Conceptually, applying Fluidic Shaping in space is simpler than doing so on Earth, as no immersion liquid is required. In addition, its scale-invariance makes it an excellent candidate for the creation of not only small components, such as ophthalmic lenses, but also of extremely large optics, such as those required for telescopes. The latter is manifested in the FLUTE project, which aims to overcome launch constraints and create a 50-meter liquid-based space telescope.[23,24] In previous work, we reported parabolic flight experiments in which we created liquid lenses under microgravity without the need for an immersion liquid.[25] We later extended this work to the creation of mirrors, using reflective liquids.[26] However, a fundamental challenge in parabolic flights is the acceleration profile, composed of repeating sequences of short (10-20 s) and noisy microgravity maneuvers. These conditions preclude lens curing, which requires an extended duration of precise microgravity. For this reason, in our parabolic flight experiments to date, all lenses and mirrors remained in liquid form.

In this paper, we report a step forward towards in-space manufacturing of optics by demonstrating the fabrication of lenses through two scientific experiments performed aboard the ISS. The experiments were performed in April 2022 as part of the Rakia Mission - the Israeli astronaut work plan aboard the ISS, as part of Axiom Space's Ax-1 mission. In the first experiment, we created centimeter-scale lenses in microgravity and cured them using UV light. The solid lenses were then returned to Earth for testing and characterization. In the second experiment, we demonstrated the scale invariance of the method by deploying a 172 mm diameter lens made of water. We then performed image analysis on the video footage of the experiment to estimate the quality of the resulting lens.

## 2. Methods

### 2.1. Experimental Hardware

Figure 2(a-d) presents the experimental hardware used to fabricate lenses under microgravity on the ISS. While Fluidic Shaping is material-agnostic, and any type of polymer, including two-component polymers and photopolymers, can be used, we opted for UV-curable polymers as they do not require bubble-free mixing and allow the astronaut to monitor and correct the injection before initiating the curing process. We thus designed the lens manufacturing chamber to serve both as a mechanical platform in which the liquid lenses were created, and as a UV chamber, in which the lenses were subsequently solidified. Figure 2(a-b) shows isometric and cross-section CAD

views of the chamber, which was composed of a base and a hinged lid. At the heart of the lens manufacturing chamber is a bounding frame with an internal acrylic ring, used as the pinning boundary for the Fluidic Shaping process. Under microgravity, upon the injection of liquid into the frame, a spherical lens is formed, as depicted in Figure 1 and detailed in Section 1 of the supplementary information. To allow both convenient access during the injection and unobstructed UV exposure during curing, the bounding frame was supported in the center of the chamber by a fixed structure (the 'frame holder'), such that the surfaces of the lens would be exposed from all directions. Two wing screws were used to affix the bounding frame to the frame holder, allowing to quickly place and remove bounding frames from the holder. The frame holder was attached to the base of the chamber using four standoffs. Two UV light-emitting diode (LED) boards were attached to the base and lid, at equal distances below and above the bounding frame, and were covered with an optical diffuser made from a layer of parchment paper. The internal dimensions of the chamber and the distances between the bounding frame, LED boards, and diffusers were experimentally optimized to produce uniform UV illumination during the curing process.

Each UV LED board consisted of two arrays of 365 nm surface-mount technology LEDs: an array of 16 low-power LEDs (OCU-400 UB365, OSA Opto Light GmbH, Germany) powered at 15 mA, and an array of two high-power LEDs (ATDS3534UV365B, Kingbright, Taiwan) powered at 360 mA. The currents through each LED were regulated using two-terminal constant LED drivers (AL5809, Diodes Inc, USA). To provide sufficient heat dissipation from the LED boards, the back surface of the board was exposed to the environment through holes in the enclosure. Both LED boards were connected through a switch (not shown in the figure) to a box with eight AA alkaline batteries connected in series. The switch allowed the astronaut to separately turn on either the low- or high-power arrays. To adhere to the ISS safety regulations, the length of the battery power line was minimized by connecting the switch to the LEDs via a relay box, which was placed right next to the battery. In addition, a limit switch was placed on the lid of the chamber to prevent the LEDs from working while the lid is open, and all the external surfaces of the chamber were coated with Kapton tape to prevent any exposure to UV, even when the lid is closed.

Three different photopolymer types were used as lens materials: a polyurethan-based resin (VidaRosa J-2D-UVDJ250G, Dongguan, China) and two optical adhesives (NOA 61 and NOA 63, Norland Products Inc., NJ, USA). These materials were selected based on our previous experience in lens fabrication in the lab. Each material was filled into a Luer-lock syringe, which was capped and wrapped in Kapton tape to keep the polymer protected from ambient UV during handling and transport. The Kapton-coated syringes were then packed in UV-opaque bags and, in contrast to the rest of the equipment which was delivered to the ISS several months in advance on a resupply mission, the polymers were launched just prior to the experiment together with the Ax-1 crew.

The UV curing experiment was performed inside the Life Science Glovebox (LSG), located in the Japanese *Kibō* module of the ISS. The LSG kept the experiment in negative pressure and thus protected the crew from possible outgassing or spillage of the liquid polymers. During the experiment, the astronaut connected the lens manufacturing chamber to the workbench of the LSG using four Velcro pads, which were placed under the base of the chamber. To create the lens, the astronaut opened the lid, placed an empty lens frame into the frame holder, and injected liquid polymer into the acrylic ring to create the lens. Once the liquid stabilized, the astronaut closed the lid, turned on the low-power UV arrays for 4 min, and then turned on the high-power UV array for another 4 min to finish the curing. After the curing, the astronaut opened the lid, removed the solidified lens, stowed it in a dedicated box, and repeated the process with a new bounding frame. The experiment was monitored and recorded using three cameras: a GoPro camera providing side view, an over-the-shoulder Canon XF705 camcorder, and a Nikon D-6 outside of the LSG.

Figure 2(e-f) shows isometric and cross-section CAD views of the frame used for the second experiment, where a large-scale (172 mm in diameter) liquid lens was deployed in the open cabin. The only liquid that could be used

safely outside the glovebox is water, and thus, this experiment was based entirely on the standard water supply of the ISS. The main challenge with deploying large-scale lenses is creating a stable and continuous fluid connection across the entire aperture while maintaining precise pinning boundary conditions. For this reason, we capped the frame with a flat acrylic window so that the liquid would wet the entire aperture, resulting in a plano-spherical lens. The acrylic window was covered with an annular PTFE plate, into which a thin acrylic ring with a sharp edge was fitted. The acrylic window and ring formed a shallow circular dish, which provided the boundary conditions (through adhesion to the window surface and pinning to the ring edge), while the PTFE was used as a hydrophobic layer to prevent water from advancing and wetting the entire plate. At one end of the frame, an aluminum mount with a ¼'' threaded hole was installed to provide an interface to a Bogen arm – a standard ISS attachment boom.

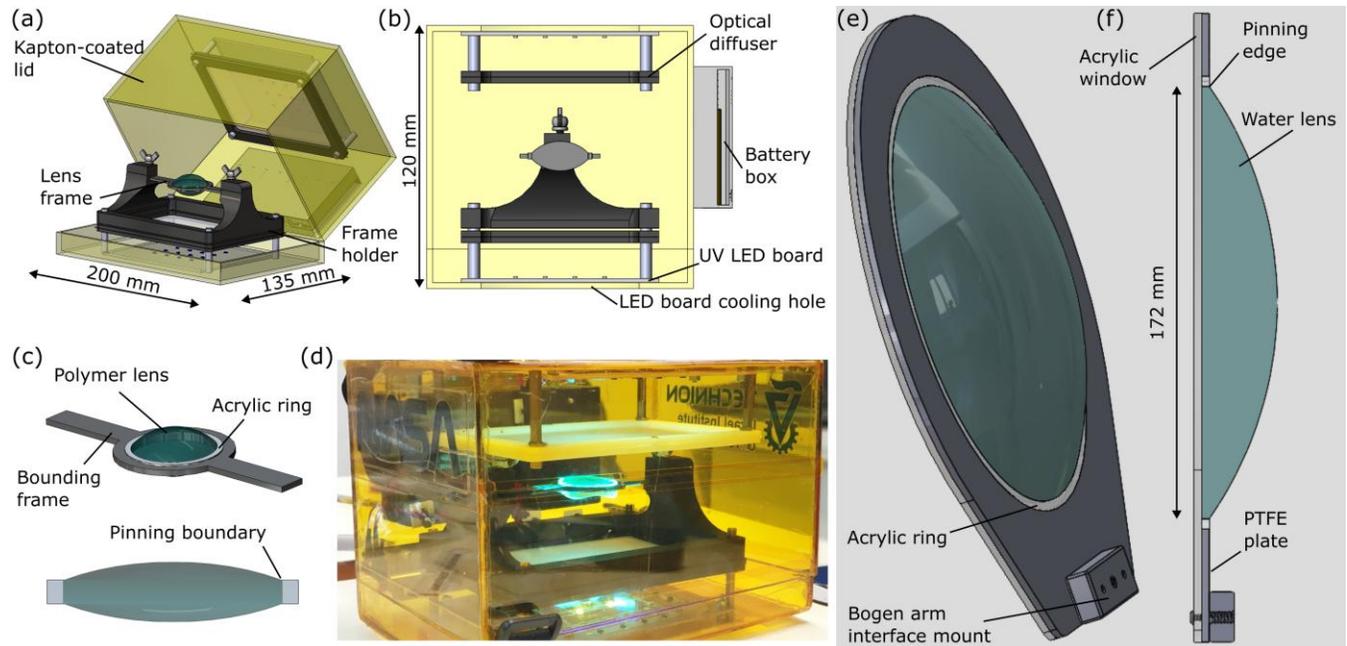

**Figure 2. The hardware developed for the creation of solid and liquid lenses on the ISS.** (a) Isometric and (b) cross-sectional views of the lens manufacturing chamber, which was used to shape and solidify polymer lenses. The chamber is composed of a base and a hinged lid, forming a closed chamber. Liquid polymer was injected into the lens frame, which was fixed at the center of the chamber by the frame holder. Once the lens was formed, the lid was closed, and UV LED boards at the base and on the lid of the chamber were turned on to polymerize the lens into a solid component. An optical diffuser was placed between the LEDs and the lens to provide uniform illumination. (c) The lens frame, into which the polymer is injected, includes an inner acrylic ring that provides pinning boundary conditions for the liquid to form the lens, and two elongated beams that secure it to the frame holder through two wing screws. (d) An image of the chamber, as taken on Earth prior to launch, with an empty bounding frame and the LEDs turned on. (e) Isometric and (f) cross-section views of the frame used to deploy a 172 mm diameter plano-spherical water lens. The frame was made of a flat acrylic window, attached to an annular PTFE plate with a thin acrylic ring fitted into it, which formed a shallow circular dish. An interface mount with a ¼'' hole was connected to the bottom part of the frame for interfacing with standard Bogen arms available on the ISS.

## 2.2. Astronaut Training

Before the Ax-1 mission launched, several training sessions were held to get both of Ax-1 astronauts who were in charge of the experiment familiar with creating lenses using Fluidic Shaping, as well as with operating the specific experimental hardware. Figure 3 shows select images from our training sessions in Axiom Space's facility in Houston, Texas. Figure 3(a-b) shows the setup that was used to train the astronauts on injecting and curing polymer lenses. The training setup consisted of a lens frame and frame holder identical to the ones used in the flight unit. To simulate microgravity conditions, the frame was submerged in a mixture of water and glycerol at a density equal to that of the polymer, such that neutral buoyancy was achieved. A standard commercial off-the-shelf mercury lamp (DR301-C, Melody Susie, US) was placed on top of the liquid immersion container and was used as a UV source to cure the lenses. Figure 3(c-d) shows the astronauts injecting a liquid photopolymer into the lens frame under neutral buoyancy. The astronauts repeated the experiment several times and practiced achieving proper pinning on the frame's edges, removing bubbles, and curing. Figure 3(e) shows astronaut Eytan Stibbe holding such a solidified lens, immediately after taking it out of the immersion liquid. Figure 3(f) shows both astronauts examining the frames intended for the deployment of large-scale water lenses.

An additional training session was held in the simulator of the ISS LSG at NASA Johnson Space Center (JSC) in Houston, Texas. The objective was to familiarize the astronauts with the LSG, as well as plan the exact positioning of the experimental hardware inside the LSG.

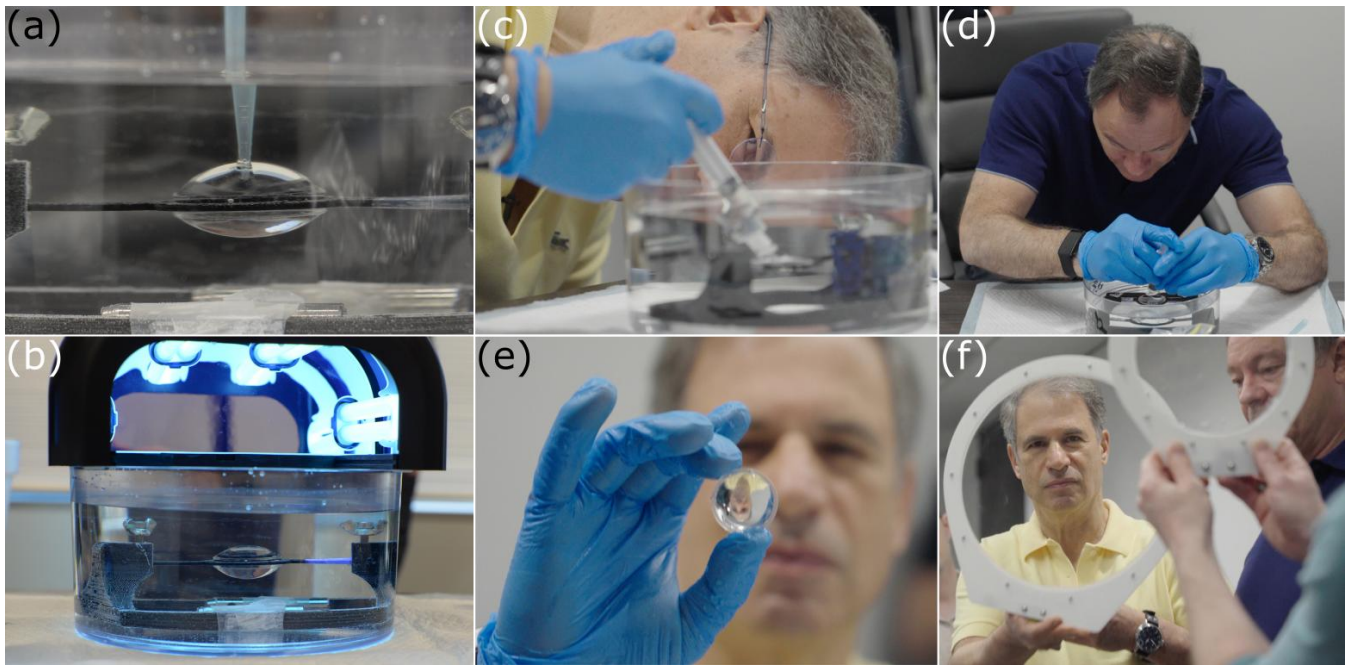

**Figure 3. Astronaut training sessions prior to Ax-1 launch.** (a-b) The training setup, based on neutral buoyancy to simulate microgravity. A replica of the frame holder was placed at the bottom of a tank full of an immersion liquid, and liquid polymer was injected into the bounding frame to create the lens. A UV lamp was used to cure the lenses. (c-d) Astronauts Eytan Stibbe and Michael López-Alegría training in creating lenses under neutral buoyancy. (e) One of the solidified lenses produced during the training session after removing it from the immersion liquid. (f) The astronauts examining the frames intended for deploying large-scale lenses.

## 3. Results and Discussion

### 3.1. UV curing of polymer lenses

The UV curing experiment was conducted in the LSG over the course of three hours, during which a total of four polymer lenses were fabricated. Two lenses were made with VidaRosa, one lens was made with NOA 61, and the last one was made with NOA 63. Figure 4(a-c) presents select images from the experiment, showing filling of the frame to create the lens, and a side view of the liquid lens prior to curing. Figure 4(d) shows the astronaut investigating the first solid lens after curing. All four lenses were packed by the astronaut in a dedicated box and returned to Earth in a subsequent cargo resupply mission for inspection in the lab.

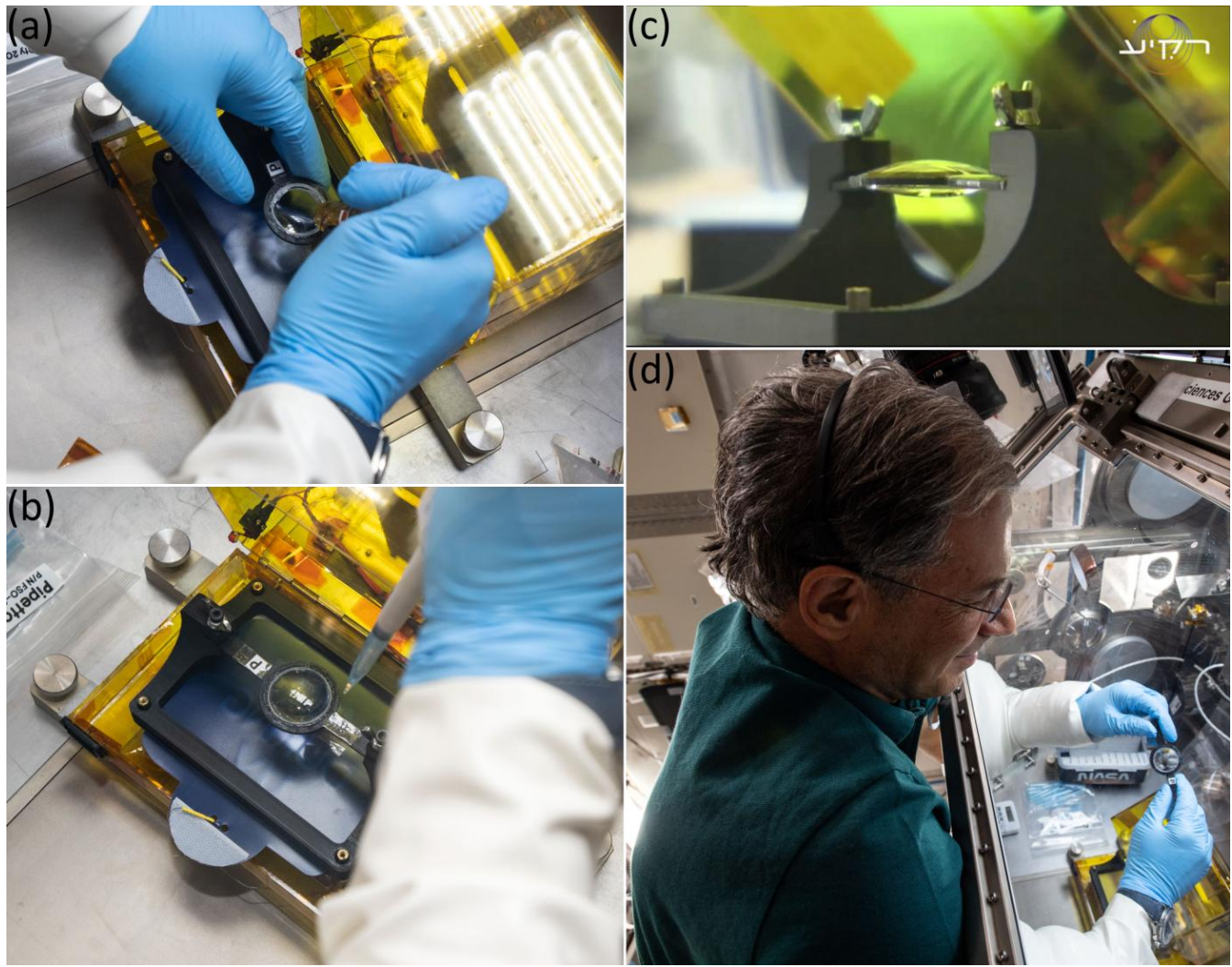

**Figure 4. In-space manufacturing of polymer lenses using Fluidic Shaping in the LSG.** (a) Filling the frame with liquid polymer under microgravity, in the same fashion as was done in the training (Figure 3), but without the need for an immersion liquid. (b) Once the frame was filled, a liquid lens was obtained. (c) A side view showing the lens before curing. (d) The astronaut holding the solidified lens after curing. Photographed by NASA astronaut Kayla Barron.

The first two lenses were fabricated from VidaRosa. After curing the first one, the astronaut visually inspected the solidified lens and noticed many small dimples across the surface of the lens. After observing the same phenomenon in the second lens, we decided to fabricate the third and fourth lenses from NOA 61 and 63, respectively. These did not exhibit such defects.

Figure 5 shows lab characterization results of two of the lenses - a VidaRosa lens and a NOA 61 lens, both with a diameter of 30 mm. Figure 5(a) shows a photograph of a VidaRosa lens, and Figure 5(b) shows a magnified view of the dimples on the surface. The dimples are in the range of 0.1-1 mm in diameter, and their appearance suggests bubble formation on the surface of the lens, which might be a result of some form of boiling of the polymer components during the curing process, either due to excess temperature rise and/or as a product of the reaction itself. This is supported by Video 2 of the Supplementary Information, which shows fumes escaping the chamber as the astronaut opened the chamber's lid after polymerization. Figure 5(c-d) shows topographical mapping of the central part of the lens surface, performed using a digital holographic microscope (DHM R1000 with 20x objective, LynceeTec, Switzerland). Due to the large sag of the lens, we were able to scan only part of the aperture with moderate slopes, as steeper slopes could not be captured by the instrument. Although the dimples clearly distort the surface, the base shape nominally forms a convex spherical cap, with a radius of curvature of 33.6 mm and an asphericity of 14.61 µm RMS. Figure 5(e) shows a nanoscale topographical mapping performed with an atomic force microscope (AFM, Asylum MFP3D, Olympus AC240 cantilever operated in tapping mode). The dimples do not appear to affect the nanoscale topography, and the surface roughness is within sub-nm RMS values ($R_q$ = 0.83 nm, over a 8x8 µm area), consistent with previous results obtained via Fluidic Shaping.[21,22] Figure 5(f-j) shows a similar characterization for the NOA 61 lens. Except for several bubbles close to the free surface, the lens is free of local imperfections. Figure 5(g) shows an image of a grid pattern taken through the lens, showing the expected pincushion distortion due to its bi-convex morphology. As shown in Figure 5(h-i), topographical mapping confirms that the NOA 61 lens is much closer to a perfect spherical topography than the VidaRosa lens, with asphericity of 6.33 µm RMS. AFM measurements also confirm sub-nm surface roughness for the NOA 61 lens ($R_q$ = 0.91 nm, over a 10x10 µm area).

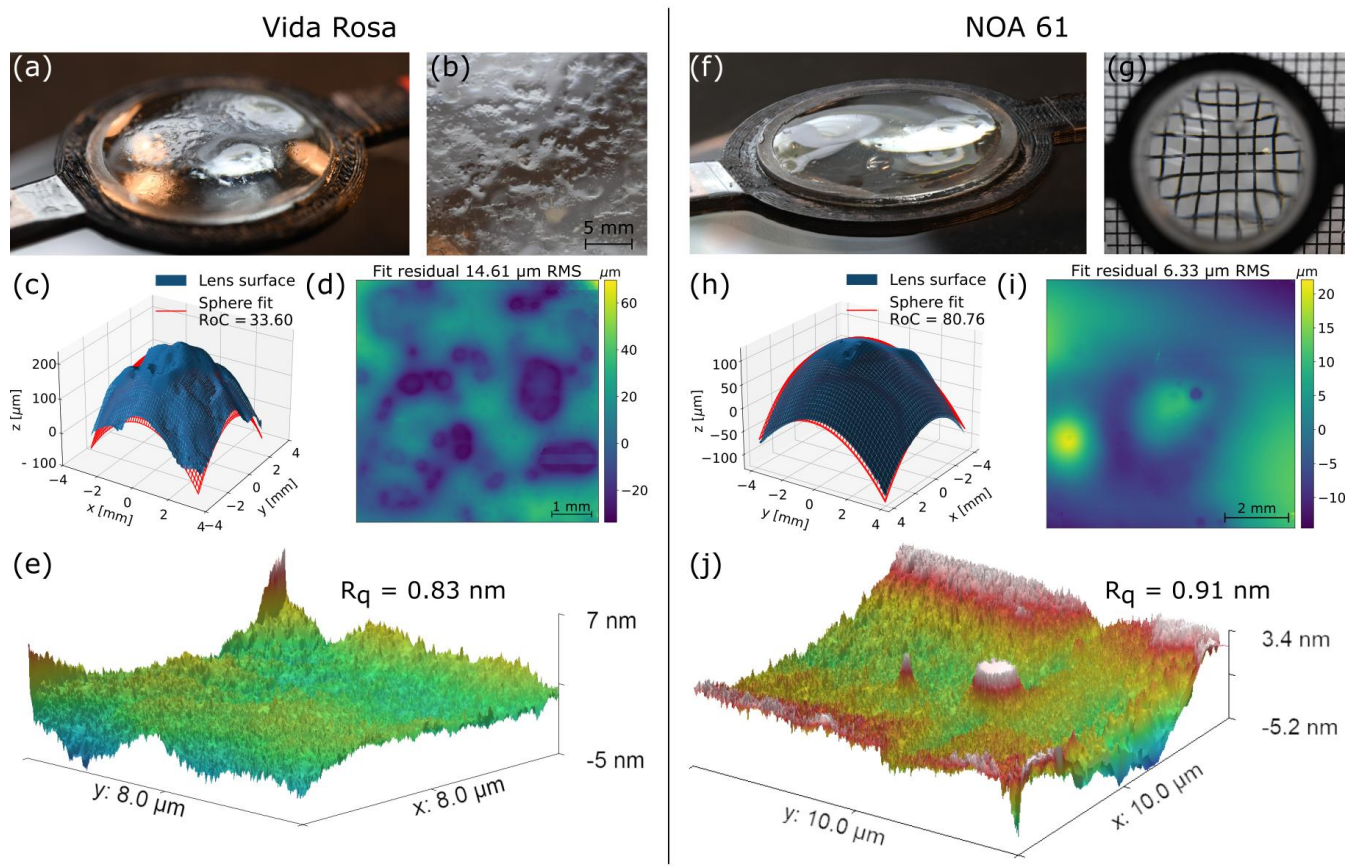

**Figure 5. Lab characterization of the cured Vida Rosa and NOA61 lenses.** (a,f) Photographs of the lens fabricated in space. (b) Close-up view showing the dimples on the surface of the Vida Rosa lens. (g) A photograph of a grid pattern taken through the NOA61 lens to demonstrate its refractive ability. (c,h) Topography maps around the center of the top surfaces of the lenses, showing the surfaces and their best-fit spherical caps. (d,i) Deviation of the surfaces from perfect spheres, obtained by subtracting the sphere fits from the measured surfaces. (e,j) AFM mapping of the surface, showing sub-nm surface roughness values for both lenses.

## 3.2. Deployment of a large-scale liquid lens

Figure 6 shows the deployment of a large-scale liquid lens in the US *Destiny* module of the ISS. The size of the setup required this experiment to be performed in the open cabin. A camera (XF705 camcorder, Canon, Japan) was positioned on a Bogen arm, facing the starboard wall. In front of the camera, the frame was fixed to another Bogen arm, such that the frame was parallel to the wall, sufficiently far from the air vents to prevent disturbances due to air flow. A 200 ml syringe was filled with water, taken directly from the drinking system of the station. The astronaut deployed the lens by injecting water directly onto the frame's acrylic window, and another astronaut refilled the syringe each time it depleted. Figure 6(a-d) shows the process of filling the frame to create the lens. During the injection, the astronaut moved the syringe in a spiral motion to properly pin the liquid to the edge of the acrylic ring. Once all the liquid fronts met and a continuous volume filled the frame, a spherical plano-convex lens was formed, as shown in Figure 6(e). This technique required a relatively large initial volume of water, resulting in a liquid lens with an exaggerated curvature. The manual process also resulted in air bubbles trapped within the lens. Since the lens remained liquid (and thus self-healing), the astronaut could aspirate the liquid with a syringe to remove the bubbles and, at the same time, reduce the lens' curvature. Once the lens was brought to the desired curvature, its magnification was demonstrated by standing near its focal plane, as demonstrated in Figure 6(f-g). The astronaut further aspirated the liquid volume several times, with apparent real-time changes to the image magnification, as shown in the Supplementary Information Video 3. After the experiment had finished, the astronaut aspirated the water back into the syringe and the fed it back to the water system of the ISS for recycling. Remaining water was absorbed with a towel.

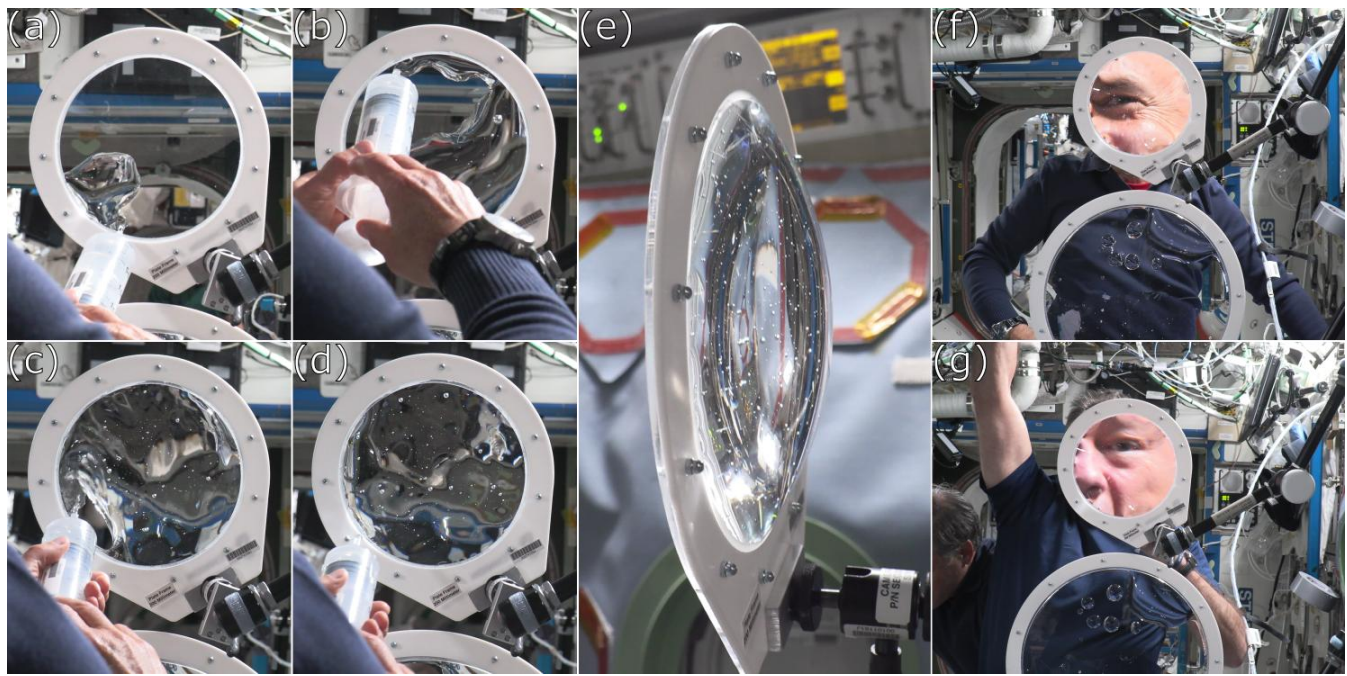

Figure 6. Deployment of a large-scale liquid lens. (a-c) Filling of the frame in a spiral motion to properly pin the water to the acrylic edge and adhere to the acrylic window. (d) Once the frame was filled, a plano-convex lens was formed. (e) A side view showing the stable lens after the liquid has settled. (f-g) Astronauts Eytan Stibble and Michael López-Alegría demonstrating the magnifying effect of the lens. Photographed by Ax-1 astronauts Michael López-Alegría and Eytan Stibble.

The experiment was originally intended solely as a qualitative demonstration. Nevertheless, the video footage can also be used to provide quantitative estimates for the lenses' performance. A simple and universal quantity for such an estimation is the modulation transfer function (MTF), which represents the resolving ability of an optical system, i.e., its ability to transfer spatial frequencies from the object space to the image space.[27,28] Among the many different ways to measure the MTF, the most natural one to apply in our case is the *slanted-edge method*, which was standardized as part of ISO 12233.[27,29–32] We refer the reader to the Standard for a full explanation of the method, and here provide only a brief description of its implementation. In short, a sharp edge target (a one-dimensional step function in intensity) is placed at the object plane, and the image of that edge is then used to extract the inherent frequency response of the system. The sharp edge is purposely positioned at a slant angle to the pixel grid of the sensor, to allow for super-sampling of the edge's position, to within sub-pixel precision.

A general difficulty in interpreting MTF measurements is that the frequency response is a property of an entire optical system rather than of a specific element within it. In our case, the contrast of the obtained image is a function of the sharpness of the target, the design and quality of both the liquid lens and the camera lens, the performance of the image sensor (including any digital manipulation performed by the camera processor), and the exact configuration in which all the components are positioned with respect to one another. Therefore, to associate the performance metrics of the system with an individual element, its contribution to the measured response must be dominant. In the case of this experiment, we made the assumption that — given the high quality of the rest of the elements of the optical chain — the liquid lens is the dominant contributor to the degradation in image sharpness. To validate this assumption, we set up a numerical Zemax model wherein the liquid lens is modeled as a plano-spherical lens, and the rest of the optical system and sensor are ideal. Agreement between the measurements and simulation would cross-validate the two and confirm that the liquid lens is indeed the central source of aberrations.

Figure 7(a) shows the selected image for the analysis, where the Rakia mission patch is positioned behind the lens and produces a magnified image. The white text on the dark background creates a slanted edge appropriate for the MTF analysis. As the experiment was not originally designed to be quantitative, the positions of the different elements, along with the focal length and aperture settings of the camera, were not precisely defined, nor were they recorded during the experiment. Moreover, some of these parameters, such as the camera position and focus setting, have changed during the experiment. For this reason, we must retrieve these parameters via triangulation, based on known dimensions of the elements in the ISS module. Figure 7(b) shows a schematic drawing of the experiment setup. The camera lens (with an effective focal length of $f_c$) and the liquid lens are positioned at a distance $l_c$ from each other, and at a lateral distance $l_d$ between their optical axes. The target is positioned at a distance $l_t$ behind the lens and decentered by a distance $l_p$ from its optical axis. The longitudinal distance between the liquid lens and the starboard wall is denoted by $l_w$, and is estimated to be 2029 mm from CAD drawings of the Destiny module, provided by NASA.

The first step of the analysis is to estimate the distance $l_c$ and the focal length of the camera $f_c$. Consider an arbitrary point $W$ on the starboard wall, which is imaged on the camera sensor. A ray transmitted from that point intersects the plane of the liquid lens at point $C$ and continues toward the camera lens. For the estimation of $l_c$ and $f_c$, we neglect the physical size of the camera lens because it is negligible compared to the longitudinal distances between the camera, liquid lens, and starboard wall, which are several meters each. We denote the scale of the image (in mm/pixel) on the planes of the starboard wall and the liquid lens as $p_w$ and $p_c$, respectively. We use features small enough (in terms of angular size in the image) and close enough to the optical axis of the camera, so that both parallax and distortion errors are negligible, and thus consider $p_w, p_c$ to be uniform on each of the planes. On the lens' plane, the lens frame is the reference feature, with a diameter of 172 mm. On the starboard wall plane, the internal ISS intercom device (Audio Terminal Unit, or ATU) serves as the reference feature, with a height of 10.5" as shown in Figure 7(d) (data kindly provided by NASA). The lengths $l_c$ and $l_w$ can be related to the lateral distances $x_w$ and $x_c$ between the camera optical axis and the points $W$ and $C$, respectively, by

$$l_c = \frac{1}{x_w / x_c - 1} l_w \tag{1}$$

Since points $W$ and $C$ are imaged on the exact same point on the sensor, $x_w$ and $x_c$ have the same size in pixels, and so the ratio $x_w / x_c$ is simply the ratio of the scale of the image in the two planes,

$$\frac{x_c}{x_w} = \frac{p_c}{p_w} \tag{2}$$

Substituting equation (2) into equation (1), and using the known geometrical parameters, we obtain $l_c = 1186 \text{ mm}$. The focal length $f_c$ is determined using geometric optics by equating the observed and theoretical magnification. The magnification $M$ is defined as the ratio of image to object size, and can be estimated from the image as $M = \frac{d_s}{N_c p_c}$, where $N_c$ is the width of the image in pixels and $d_s$ is the width of the sensor in mm. For a thin lens, the theoretical magnification is given by $M = \frac{f_c}{l_c - f_c}$ [28]. Equating the two expressions and solving for $f_c$ yields

$$f_c = \frac{d_s}{p_c N_c + d_s} l_c. \tag{3}$$

For $N_c = 3840 \text{ pix}$ and $d_s = 13.2 \text{ mm}$ (provided in the manufacturer's specification sheet), we obtain an estimate of $f_c \approx 19 \text{ mm}$, which is within the range of the zoom camera lens (8.3 – 124.5 mm).[33] In contrast to the focal length, the size of the camera lens aperture cannot be directly extracted from the image. For this reason, we can only bound the entrance pupil size between the two extreme values of $f/2.8$ and $f/27$, according to the camera specifications.[33]

The remaining parameters required to fully retrieve the optical setup are the longitudinal and lateral distances of the patch relative to the axis of the liquid lens, $l_t$ and $l_p$. We determine their values using Zemax, by constructing the optical model and optimizing the distances to produce the same image size observed in Figure 7(a). To that end, we model the liquid lens as a plano-convex lens with a refractive index of 1.33[34] and a radius of curvature of 144 mm, determined by a best-fit spherical cap as shown in Figure 7(c). Since the camera is proprietary and the internal structure of its compound lens is not known, we model it as a single paraxial surface with a focal length of $f_c$, and a diameter equal to the entrance pupil in the two possible scenarios (either $f_c/2.8$ or $f_c/27$). The full parameters of the resulting Zemax model are available in Figure S2 in the Supplementary Information.

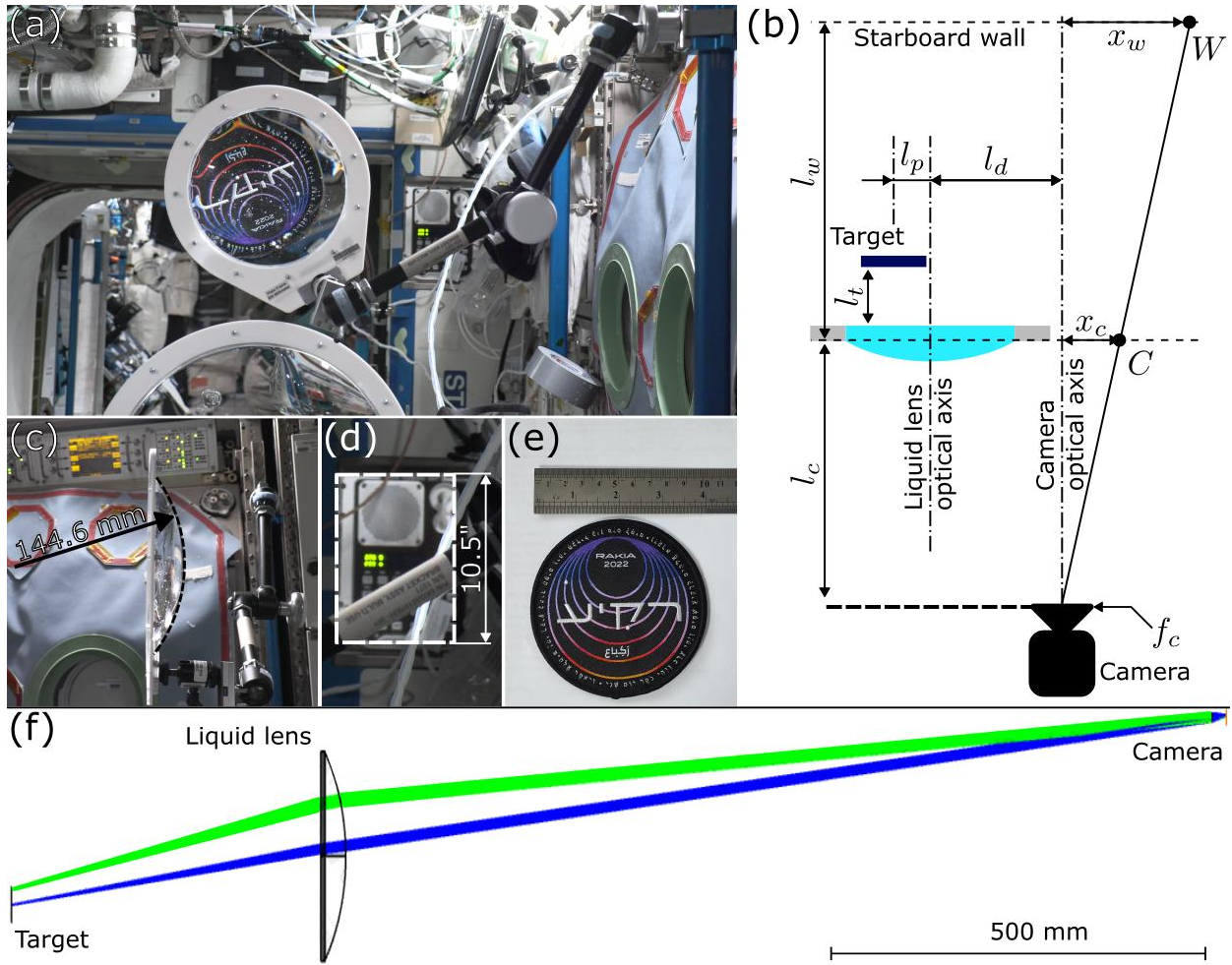

**Figure 7. Retrieval of the experimental parameters through triangulation on the image.** (a) The chosen image from the video footage, showing the Rakia mission patch magnified by the liquid lens. The image was chosen such that the details on the mission patch are as sharp as possible and oriented so that the slant angle is suitable for quantitative MTF analysis. (b) A schematic drawing of the experimental setup (not to scale). The target was placed behind the liquid lens, and the magnified image was captured by the camera. Using the dimensions of objects observed in the image and the distance between the planes of the liquid lens and the starboard wall, the distance between the camera and the liquid lens, as well as the effective focal length of the camera lens, are estimated via triangulation. (c) A side view of the liquid lens, showing curve fitting of the optical surface to a spherical cap, providing an estimate for its radius of curvature. (d) The internal intercom system of the ISS, used to estimate the scale of the image at the starboard wall plane. (e) A photograph of the Rakia mission patch, with a ruler for scale. (f) After the distance between the camera and liquid lens is obtained, the optical train is modelled in Zemax, with $l_t$ and $l_p$ as free parameters, which are varied until the patch image matches the magnification and decenter in the experiment. The camera lens was modelled as a single paraxial surface with a diameter equal to the entrance pupil. Photographed by Ax-1 astronaut Eytan Stibbe.

Once the optical setup parameters are retrieved, we are able to move on to the MTF analysis on the image. The image, taken from a video, is sRGB and so the analysis is performed on each color channel (red, green and blue) separately. We first perform gamma decoding to remove the nonlinear corrections performed by the camera processor and obtain an image with intensity values that linearly depend on the incident radiant flux. The opto-electronic transfer function (specified by the manufacturer) follows the ITU-R BT.709 standard,[33,35,36] with the inverse transformation given explicitly for each pixel separately as

$$I = \begin{cases} V/4.5 & \text{if } V < 0.081 \\ \left(\dfrac{V+0.099}{1.099}\right)^{1/0.45} & \text{if } V \geq 0.081 \end{cases}, \quad (4)$$

where $V$ is the encoded color intensity as saved by the camera, and $I$ is the decoded intensity. Figure 8(a) shows the gamma-decoded version of the selected image. The inset shows a zoomed-in view of the relevant region on the target, where $i$, $j$ represent the indices for the row and column, in the pixel array, respectively. The yellow rectangle in Figure 8(b) is the region of interest (ROI) on the edge, in which the analysis is performed. For each pixel column j and color channel $k \in \{\text{Red},\text{Green},\text{Blue}\}$, we take a central difference of the intensity profile, $I\big|_{j,k}(i)$, and define the maximum of its absolute value as the position of the edge, corresponding to the inflection point of the intensity profile. We then fit a linear curve $i_E(j,k) = m(k)\cdot j + n(k)$ to those points, defining the edge line, as indicated by the dashed lines in Figure 8(b). Next, an edge spread function $\text{ESF}[s(i,j,k),k]$ is constructed as the intensity of each pixel within the ROI as a function of its normal distance $s$ from the edge,

$$s(i,j,k) = (i - i_E(j,k))\cdot \sin(\arctan(-m(k))) = (i_E(j,k) - i)\dfrac{m(k)}{m^2(k)+1}. \quad (5)$$

The slant angle provides sub-pixel precision, as pixels from different columns in the ROI are projected together onto a single intensity profile without overlapping. This results in a "super-sampled" ESF with a sample resolution higher than the original image. To remove high-frequency noise from the ESF, we perform local regression with the LOWESS algorithm[37,38], and use this smoothed version in the rest of the analysis. Figure 8(c) shows the raw and smoothed ESF as point markers and continuous curves, respectively. The ESF is essentially the step response of the system, and therefore its first spatial derivative, shown in Figure 8(d) is the one-dimensional impulse response of the system and also called the line spread function (LSF),

$$\text{LSF}(s,k) = \dfrac{\partial(\text{ESF}(s,k))}{\partial s}. \quad (6)$$

The MTF is then the modulus of the Fourier transform of the LSF,

$$\text{MTF}(\nu) = \left| \int_{\min(s)}^{\max(s)} \left(\text{LSF}(s)\cdot w(s)\cdot e^{-i2\pi\nu s}\right) ds \right|, \quad (7)$$

where $\nu$ is the spatial frequency along the $s$ direction and $w(s)$ is the Tukey window function with a shape parameter of 1.0.[30]

Figure 8(e) shows the MTF extracted from the image, in comparison with the polychromatic sagittal MTFs simulated in Zemax (dashed black curves). We compare the empirical results to polychromatic theoretical curves and not to monochromatic ones, since the color channels in the camera have wide bandwidths and spectrally overlap.[39] We do not know, however, the exact responsivity of the camera sensor, and so we use a simple average between 486 nm, 588 nm, and 656 nm, as typically done in optical designs intended for the entire visible region.

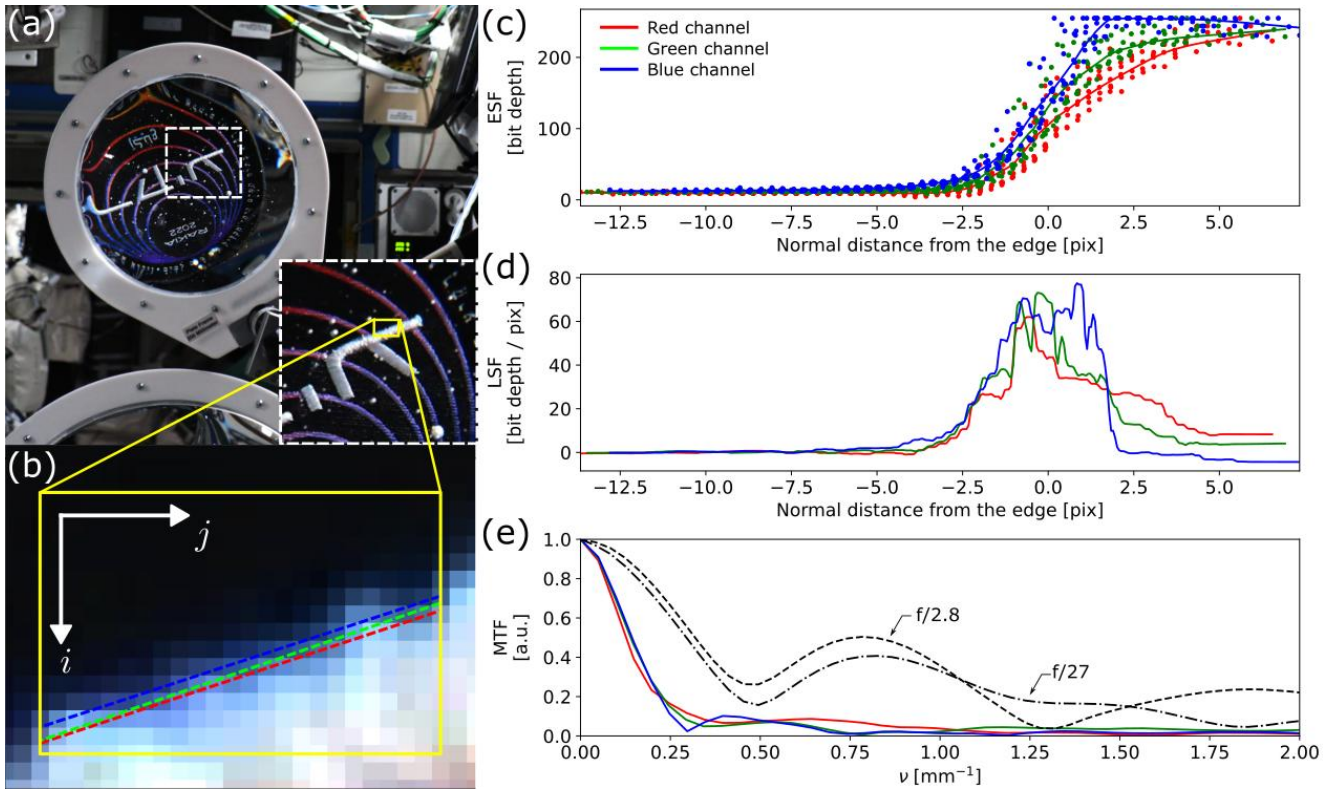

**Figure 8. Estimation of the resolving ability of the large-scale liquid lens.** (a) A close-up view of the magnified target after gamma decoding. The inset shows a zoomed-in view of the analysis area. (b) The rectangular region of interest in which the MTF analysis is performed. (c) The super-sampled edge-spread functions (ESFs) within the region of interest, and smooth curves produced via local regression. (d) The line spread functions (LSFs), computed by taking the first derivatives of the ESFs. (e) The modulation transfer functions (MTFs), which are the moduli of the Fourier transform of the LSFs, as functions of spatial frequency in the object space. The dashed and dotted-dashed black lines represent the simulated polychromatic sagittal MTF in the Zemax model, for camera lens apertures of f/2.8 and f/27, respectively. The different colors in all graphs correspond to different channels (red, green, and blue) in the sRGB image. Photograph in (a) by Ax-1 astronaut Eytan Stibbe.

We observe a reduction in modulation compared to the Zemax model, which we associate mainly with lossy video compression. Although we have used the highest-quality video format available to us (MXF format with 10 bits per channel), it is still compressed. It is impossible to precisely predict how compression affected the MTF; however, in the vast majority of cases, it leads to loss of contrast and thus to a reduction in the MTF.[40] Aside from compression, the reduction in the MTF could also be a result of optical aberrations. The most likely source of aberrations is liquid deformations on the free surface of the water lens. The microgravity levels on the ISS are known to be very precise,[41,42] yet disturbances from mechanical vibrations that coupled to the lens and may have not fully settled, as well as the air ventilation system, could be significant sources of aberration. Such aberrations can also result from imprecise boundary conditions, which lead to non-spherical geometry. Although we took extra care in the design to create ideal pinning conditions, non-perfect pinning can be observed in several points around the perimeter of the lens. This can be a result of dust particles and/or oil residues (e.g., from fingerprints). Elgarisi et al.[22] have shown that high-frequency boundary distortion exponentially decays as a function of the radial distance from the boundary. However, low frequencies are attenuated much more slowly and may still produce non-negligible aberrations throughout the entire aperture of the lens. Another possible source of optical aberrations in the liquid lens is scattering by air bubbles trapped within the water lens during deployment.

## 4. Conclusions and Outlook

In this work, we report on the first implementation of Fluidic Shaping for in-space manufacturing of optical components, in a set of experiments performed on board the ISS. Using a lens manufacturing chamber and a large-scale deployment frame designed specifically for these experiments, we demonstrated the creation and UV curing of centimeter-scale polymer lenses and the deployment and characterization of a large liquid water lens.

While lab characterization of our UV-cured lenses confirms sub-nanometric surface roughness, similar to that obtained by Fluidic Shaping in neutral buoyancy, we observed unexpected deformations on the lenses made from the VidaRosa polymer. These deformations suggest a thermo-chemical phenomenon – likely localized boiling due to heat generated during polymerization. In our setup, we deliberately chose to enclose the lens in a chamber during curing to prevent potential disturbances due to vent airflow inside the LSG. As microgravity eliminates natural convection, our enclosed chamber created a highly insulated thermal environment, where heat could dissipate from the lens only by conduction through the surrounding air and by radiation. These conditions differed markedly from those under which we optimized our system on Earth. Although we had tested equally thick polymer layers without noticeable deformation, it is likely that natural convection in those settings was sufficient to remove excess heat during curing. In contrast to the VidaRosa lenses, the NOA lenses showed no such defects, indicating that their polymerization chemistry generates less heat or follows different thermal dynamics. These findings point to an important direction for future research: investigating the kinetics and thermodynamics of polymerization under microgravity, with particular attention to heat transfer mechanisms, curing environments, and material-specific behavior.

Our large-scale liquid optics experiment demonstrated both the scale-invariant and material-agnostic nature of Fluidic Shaping. However, the manual process of spreading the liquid across a large aperture, while also achieving proper wetting and pinning, proved challenging. This highlights the need for more advanced liquid deployment strategies suitable for large apertures with minimal bubble formation. Such capabilities will be essential for future liquid-based space observatories such as FLUTE. While the experiment was primarily intended as a qualitative demonstration, we were able to extract first-order quantitative performance metrics through image analysis. To achieve higher precision and consistent optical quality, future systems will require integrated, *in-situ* characterization tools capable of monitoring fluidic optical component formation in real time, similar to what we have implemented in our parabolic flight setups.[25,26]

In both experiments, astronaut involvement provided valuable flexibility in orbit, enabling real-time response to unexpected developments. However, manual liquid handling also introduced operational challenges, such as trapped air bubbles and inconsistent wetting. Future experiments should transition toward more automated systems, with crew members serving in a supervisory capacity.

## Data Availability

Videos of the water lens experiment and the raw AFM data are available in the following repository:

https://doi.org/10.5281/zenodo.16931251.

## Code Availability

The Python code used to analyze images from the water lens video is provided in the above repository.

## Acknowledgments

This project has received funding from the European Research Council under the European Union's Horizon 2020 Research and Innovation Programme, grant agreement 10104451 (Fluidic Shaping). Views and opinions expressed are, however, those of the author(s) only and do not necessarily reflect those of the European Union or the European Research Council Executive Agency. Neither the European Union nor the granting authority can be held responsible for them. We also acknowledge financial support from the Israeli Space Agency within the Israeli Ministry of Innovation, Science and Technology, and from the Norman and Helen Asher Space Research Institute (ASRI) fund. We also thank NASA Ames Center Innovation Fund for its financial support of this project and the ISS National Lab for providing material safety certification at the NASA White Sands Test Facility. M.E. was supported by the Ramon Graduate Fellowship of the Israel Ministry of Innovation, Science and Technology.

We thank Melody Korman from the Rakia Mission for mission coordination and real-time ground support before, during, and after the experiment. We thank Christian Maender from Axiom Space for his dedication and active involvement in making this experiment possible. We thank the Rakia Mission for coordinating the experiment and creating the opportunity, and in particular, Inbal Kreiss, Eliran Raphael Hamo, and Shir Stibbe, with whom we worked closely on executing the experiment as well as Ran Livne from the Ramon foundation. We thank Mandy Clayton from NASA Marshall for writing the mission procedures of both experiments and Brandon Williams from Axiom Space for his role in the experiment integration. We also thank Matan Nice from Technion for his help during the development of the curing system, and Ofek Efraim from Technion and Jessica Koehne from NASA Ames for their contributions in measuring the topography of the cured lenses. We thank Aliza Shultzer from Technion for her help with the mission's logistics. Finally, we would like to express our gratitude to Jill Bauman, Jay Bookbinder, Thomas Berndt, Meredith Blasingame, Karen Bradford, Rhys Cheung, Jacob Cohen, Matthew Holtrust, Robert Padilla, and Harry Partridge of NASA Ames, as well as Kent Bress, Judith Carrodeguas, Trenton Roche, Brian Stanford, and Brian Wessel of NASA HQ for enabling and supporting this collaborative effort.

## Author Contributions

OL, ME, VF, MB and EB conceived the experiment. OL designed the experiment hardware, with contributions from ME, AR, SP and MB. ME and OL optimized the curing sequence. OL performed the safety analysis and prepared the review board documentation, with contributions from ME. OL, AR, ME, SP, JE, KG, DW, IG and MB assembled the experimental hardware. ME, VF, OL, MB, RB, and CC trained the astronauts. ES performed the experiments aboard the ISS, with contributions from MLA. CC supervised the experiment performance from mission control center on Earth. OL analyzed the experiment data. OL and MB wrote and edited the manuscript. ES and EB reviewed, commented, and corrected the manuscript. All authors approved the manuscript.

**Competing Interests**

The authors declare no competing financial or non-financial interests.

**References**

1. Werkheiser, M. J. *et al.* 3D Printing In Zero-G ISS Technology Demonstration. in *AIAA SPACE 2014 Conference and Exposition* (American Institute of Aeronautics and Astronautics). doi:10.2514/6.2014-4470.

2. Hoffmann, M. & Elwany, A. In-Space Additive Manufacturing: A Review. *J. Manuf. Sci. Eng.* **145**, (2022).

3. Patane, S., Joyce, E. R., Snyder, M. P. & Shestople, P. Archinaut: In-Space Manufacturing and Assembly for Next-Generation Space Habitats. in *AIAA SPACE and Astronautics Forum and Exposition* (American Institute of Aeronautics and Astronautics). doi:10.2514/6.2017-5227.

4. Sacco, E. & Moon, S. K. Additive manufacturing for space: status and promises. *Int. J. Adv. Manuf. Technol.* **105**, 4123–4146 (2019).

5. wpadmin. Made In Space Receives NASA Award to Advance Space-Based Manufacturing. *ISS National Lab* https://issnationallab.org/iss360/made-in-space-nasa-award-advance-space-based-manufacturing/ (2020).

6. Makaya, A. *et al.* Towards out of earth manufacturing: overview of the ESA materials and processes activities on manufacturing in space. *CEAS Space J.* **15**, 69–75 (2023).

7. Arnold, C., Monsees, D., Hey, J. & Schweyen, R. Surface Quality of 3D-Printed Models as a Function of Various Printing Parameters. *Materials* **12**, 1970 (2019).

8. Wang, W. M., Zanni, C. & Kobbelt, L. Improved Surface Quality in 3D Printing by Optimizing the Printing Direction. *Comput. Graph. Forum* **35**, 59–70 (2016).

9. Xing, H., Zou, B., Li, S. & Fu, X. Study on surface quality, precision and mechanical properties of 3D printed ZrO2 ceramic components by laser scanning stereolithography. *Ceram. Int.* **43**, 16340–16347 (2017).

10. Mader, T. H. *et al.* Optic Disc Edema, Globe Flattening, Choroidal Folds, and Hyperopic Shifts Observed in Astronauts after Long-duration Space Flight. *Ophthalmology* **118**, 2058–2069 (2011).

11. Lee, A. G. *et al.* Spaceflight associated neuro-ocular syndrome (SANS) and the neuro-ophthalmologic effects of microgravity: a review and an update. *Npj Microgravity* **6**, 1–10 (2020).

12. Nguyen, T. *et al.* Spaceflight Associated Neuro-ocular Syndrome (SANS) and Its Countermeasures. *Prog. Retin. Eye Res.* 101340 (2025) doi:10.1016/j.preteyeres.2025.101340.

13. Cavarroc, C., Boccaletti, A., Baudoz, P., Fusco, T. & Rouan, D. Fundamental limitations on Earth-like planet detection with extremely large telescopes. *Astron. Astrophys.* **447**, 397–403 (2006).

14. Macintosh, B. *et al.* Extreme adaptive optics for the Thirty Meter Telescope. in *Advances in Adaptive Optics II* vol. 6272 201–215 (SPIE, 2006).

15. Ganel, O. *et al.* NASA strategic astrophysics technology investments: a decade of benefits, outlook informed by the 2020 Decadal Survey. in *Advances in Optical and Mechanical Technologies for Telescopes and Instrumentation V* vol. 12188 57–70 (SPIE, 2022).

16. *Pathways to Discovery in Astronomy and Astrophysics for the 2020s*. (National Academies Press, Washington, D.C., 2023). doi:10.17226/26141.

17. Zhu, L., Li, Z., Fang, F., Huang, S. & Zhang, X. Review on fast tool servo machining of optical freeform surfaces. *Int. J. Adv. Manuf. Technol.* **95**, 2071–2092 (2018).

18. Xia, Z., Fang, F., Ahearne, E. & Tao, M. Advances in polishing of optical freeform surfaces: A review. *J. Mater. Process. Technol.* **286**, 116828 (2020).

19. Roeder, M., Guenther, T. & Zimmermann, A. Review on Fabrication Technologies for Optical Mold Inserts. *Micromachines* **10**, 233 (2019).

20. Zhu, W.-L. & Beaucamp, A. Compliant grinding and polishing: A review. *Int. J. Mach. Tools Manuf.* **158**, 103634 (2020).

21. Frumkin, V. & Bercovici, M. Fluidic shaping of optical components. *Flow* **1**, E2 (2021).

22. Elgarisi, M., Frumkin, V., Luria, O. & Bercovici, M. Fabrication of freeform optical components by fluidic shaping. *Optica* **8**, 1501 (2021).

23. Hall, L. Fluidic Telescope (FLUTE): Enabling the Next Generation of Large Space Observatories - NASA. https://www.nasa.gov/general/fluidic-telescope-flute-enabling-the-next-generation-of-large-space-observatories/.

24. Figliozzi, G. What is the Fluidic Telescope? - NASA. https://www.nasa.gov/science-research/astrophysics/what-is-the-fluidic-telescope/.

25. Luria, O. *et al.* Fluidic shaping and in-situ measurement of liquid lenses in microgravity. *Npj Microgravity* **9**, 74 (2023).

26. Luria, O. *et al.* Shaping a gallium alloy and an ionic liquid into spherical mirrors for future liquid-based telescopes—experimental setup and demonstration in parabolic flights. *J. Astron. Telesc. Instrum. Syst.* **10**, (2024).

27. Williams, C. S. & Becklund, O. A. *Introduction to the Optical Transfer Function*. (SPIE Press, Bellingham, Wash, 2002).

28. Born, M. & Wolf, E. *Principles of Optics*. (Cambridge University Press, Cambridge, United Kingdom, 2019).

29. Viallefont-Robinet, F. *et al.* Comparison of MTF measurements using edge method: towards reference data set. *Opt. Express* **26**, 33625 (2018).

30. ISO 12233 - Photography - electronic still picture imaging - resolution and spatial frequency response. (2023).

31. Xu, M., Cong, M. & Li, H. Research of on-orbit MTF measurement for the satellite sensors. in (eds. Tong, Q., Shan, J. & Zhu, B.) 915809 (Wuhan, China, 2014). doi:10.1117/12.2064148.

32. Hwang, H., Choi, Y.-W., Kwak, S., Kim, M. & Park, W. MTF assessment of high resolution satellite images using ISO 12233 slanted-edge method. in (eds. Bruzzone, L., Notarnicola, C. & Posa, F.) 710905 (Cardiff, Wales, United Kingdom, 2008). doi:10.1117/12.800055.

33. Africa, C. C. and N. Specifications & Features - Canon XF705 Professional Camcorder. *Canon Central and North Africa* https://en.canon-cna.com/video-cameras/xf705/specifications/.

34. Hale, G. M. & Querry, M. R. Optical Constants of Water in the 200-nm to 200-µm Wavelength Region. *Appl. Opt.* **12**, 555–563 (1973).

35. Rec. ITU-R BT.709-6: Parameter values for the HDTV standards for production and international programme exchange. (2015).

36. Rec. ITU-R BT.2020: Parameter values for ultra-high definition television systems for production and international programme exchange. (2015).

37. Cleveland, W. S. & Loader, C. Smoothing by Local Regression: Principles and Methods. in *Statistical Theory and Computational Aspects of Smoothing* (eds. Härdle, W. & Schimek, M. G.) 10–49 (Physica-Verlag HD, Heidelberg, 1996).

38. Josef Perktold *et al.* statsmodels/statsmodels: Release 0.14.2. Zenodo https://doi.org/10.5281/ZENODO.593847 (2024).

39. Sigernes, F. *et al.* The absolute sensitivity of digital colour cameras. *Opt. Express* **17**, 20211–20220 (2009).

40. Zhao, Y., Outif, A. & Stewart, I. Comparison of the effect of lossy compressions with the modulation transfer function. in *Medical Imaging 1999: PACS Design and Evaluation: Engineering and Clinical Issues* vol. 3662 324–334 (SPIE, 1999).

41. DeLombard, R., Hrovat, K., Kelly, E. & McPherson, K. Microgravity Environment on the International Space Station. in (AIAA). doi:10.2514/6.2004-125.

42. Kevin McPherson, Eric Kelly, & Jennifer Keller. Acceleration Environment of the International Space Station. in (AIAA, Orlando, Florida , USA, 2009). doi:https://doi.org/10.2514/6.2009-957.
Page 19 of 19